\documentclass[twocolumn,showpacs,revtex4,amsmath,amssymb]{revtex4}

\usepackage[T2A]{fontenc}
\usepackage[cp1251]{inputenc}
\usepackage[russian]{babel}

\begin{document}

\title{Interference effects in the reradiation of ultrashort electromagnetic pulses}

\author{V.I. Matveev}
 \email{vim1598@gmail.com}
\affiliation{Northern (Arctic) Federal University, 163002 Arkhangelsk, Russian Federation}%
\author{D.U. Matrasulov}
 \email{dmatrasulov@gmail.com}
\affiliation{Turin Polytechnic University in Tashkent, 17 Niyazov Str., 100095, Tashkent, Uzbekistan}


\begin{abstract}
Reradiation of a spatially non-uniform ultrashort electromagnetic pulse interacting with the linear chain of multielectron atoms
is studied in the framework of sudden perturbation approximation.
Angular distributions of the reradiation spectrum for arbitrary number of atoms are obtained.
It is shown that interference effects for the photon radiation amplitudes lead to appearing of "diffraction" maximums. The obtained results can be  extended to the case of two- and three-dimensional crystal lattices and atomic chains.  The approach developed allows also to take into account  thermal vibrations of the lattice atoms.\end{abstract}

\pacs{03.65.-w; 42.50.Hz; 32.80.Qk; 78.47.J}
\maketitle

Interaction of short and ultrashort electromagnetic pulses with atomic systems and discrete structures such as crystal lattices, atomic and molecular chains  is of importance both from practical and fundamental viewpoints.
Such processes can be used in X-ray structure analysis and  attosecond spectroscopy increasing their resolution.
Attosecond physics uses as the targets relatively simple systems and their similar extensions ~\cite{Agostini2004}-\cite{Krausz2009} in the context of ultrashort processes  ~\cite{Jeltikov2009}. During recent years one can observe rapidly growing interest to the interactions of ultrashort pulses which is caused by their potential applications in supestrong lasers and heavy ions accelerators where strong fields of highly charged ultrarelatuvistic ions are comparable with those of light  field. In particular, in the collision experiments of with fast highly ions the field of ultrarelativistic uranium ion,  $U^{92+}$ with the collision energy 1 GeV/nucleon can be considered as superintense  pulse with the strength $(I>10^{19}$ Wt/cm$^2$)and the duration $\tau \sim 10^{-18}$ s. Despite certain progress made in the study of interactions of ultrashort pulses with atomic systems,
reradiation induced by such interaction is still remaining as less-explored topic which is considered only few papers
(see, e.g. ~\cite{Astapenko2011} and references theirein).

Usually diffraction of X-rays on different periodic structures is described as the plane wave scattering with infinite time duration  ~\cite{Kauli}.
In ~\cite{Golovinkii2006} the scattering of ultrashort pulse on the atom is studied within the classical approach. The scattering of ultrashort electromagnetic pulse in multielectron atoms within the quantum mechanical approach is considered in the Refs.  ~\cite{{Astapenko2011},{Astapenko2010}}.

In particular, in ~\cite{{Astapenko2011},{Astapenko2010}} the spectra of scattered  radiation are obtained for different values of the pulse duration. Such an approach can be applied for attosecond pulses, too. However, exact treatment for such ultrashort pulses is possible within the sudden perturbation approach which enables also the treatment of the re-scattering processes  ~\cite{Matv2003} and can be extended to the cases of simplest molecules ~\cite{Eseev2011}.

Below we consider ultrashort pulses whose duration,  $\tau$ is much shorter that the characteristic atomic period $\tau_a$ for the Bohr orbit.

In this paper using the approach developed in the Ref.  ~\cite{Matv2003}, we treat reradiation of the ultrashort pulses of the electromagnetic field in the liner chain of the multielectron atoms. The method developed allows exact account of the non-uniform distribution of both ultrashort pulse as well as photon  momentum for reradiation processes. The field of the ultrashort pulse can be taken into account within the sudden perturbation, while photon radiation is described by the usual perturbation theory. Also we calculated angular distributions of the reradiation spectra for arbitrary number of atoms in the chain. It is shown that  interference effects in the amplitudes of the radiated photons may lead to appearing of "diffraction" maximums which become infinitely narrow in the limit of infinitely large number of atoms.

Consider an atom interacting with the electromagnetic pulse of Gaussian shape.
 (atomic units are used throughout the paper):
\begin{eqnarray}
{\bf E}({\bf r},t)={\bf E}_0 \frac{\alpha}{\sqrt{\pi}}e^{-\alpha^2(t-{\bf n}_0{\bf r}/c)^2}\cos(\omega_0 t-{\bf k}_0{\bf r})\;,
\label{iz1a}
\end{eqnarray}
 where ${\bf k}_0=(\omega_0/c){\bf n}_0$, а ${\bf n}_0$ is the unit vector directed along the pulse direction , ${\bf r}$ is the coordinate of the observation point, $c$ is the light speed and $\tau \sim 1/\alpha$ is the pulse duration. We note that  ${\bf E}({\bf r},t) \to {\bf E}_0 \delta (t-{\bf n}_0{\bf r}/c)$ for $\alpha \to \infty$.

 According to the Ref. ~\cite{Matv2003} the interaction of the atomic electrons with electromagnetic pulse can be described by the potential
\begin{eqnarray}
 V(t) \equiv V(\{{\bf r}_e\},t)=
\sum\limits_{e=1}^{e=N_e}  {\bf E}({\bf r}_e,t){\bf r}_{e} \;,
\label{iz2a}
\end{eqnarray}
with $\{{\bf r}_{e}\}$ being the collective coordinate of the atomic electrons
and $(e=1,...,N_e)$, $N_e$ is their number.

Let the value of the quantity $\alpha$ in Eq. (~\ref{iz1a}) such that $V(t)$ is effectively non-zero only during the time,
 $ \tau \sim \alpha^{-1}$,
which is much shorter that the characteristic periods of atomic electrons whose dynamics is described by the Hamiltonian $H_0$.

Then the amplitude for transition of the atom from the initial , $ \varphi _{0} $ top a final state, $ \varphi _{n} $ caused by the interaction with sudden perturbation,
$V(t)$ can be written as ~\cite{DykhneYudin1978}:
\begin{equation}
a_{0n}=\langle \varphi _{n}\mid exp(-i\int\limits_{-\infty}^{+\infty}
  V(t)dt)\mid \varphi _{0}\rangle,
\label{iz2}
\end{equation}
where $ \varphi _{0} $  and $ \varphi _{n} $  are the  eigenfunctions of the unperturbed Hamiltonian, $H_0$.
Perturbation potential given by Eq. (~\ref{iz2a}) is written for the atom whose nucleus is at the origin of coordinate.
For the atom with the nucleus at the distance  ${\bf d}$ from the origin of coordinate system the interaction potential can be written as (~\ref{iz2a})
\begin{eqnarray}
 V(t) =
 \sum\limits_{e=1}^{e=N_e} {\bf E}({\bf r}_e+{\bf d},t)({\bf r}_e+{\bf d})=
 \nonumber\\
 =\sum\limits_{e=1}^{e=N_e} {\bf E}({\bf r}_e+{\bf d},t){\bf r}_e+
 {\bf E}({\bf r}_e+{\bf d},t){\bf d} \;,
\label{iz2ad}
\end{eqnarray}
where ${\bf r}_e$ are the distance between the atomic electrons and nucleus.
It follows from Eq.(\ref{iz1a}) that for  ${\bf E}({\bf r}_e+{\bf d},t)$ the integral  $\int\limits_{-\infty}^{+\infty}
 dt$ from second term in the right hand side of Eq.(\ref{iz2ad}) ) doesn't depend on the coordinates of atomic electrons.
Therefore it doesn't make contribution into the transition amplitude, $a_{0n}$.

Thus within the sudden perturbation approach as the interaction potential one can use
\begin{eqnarray}
 V(t) \equiv V({\bf r}_e,t)=
 \sum\limits_{e=1}^{e=N_e}{\bf E}({\bf r}_e+{\bf d},t){\bf r}_e \;.
\label{iz2adf}
\end{eqnarray}

Now consider one-dimensional chain of  $N$ noninteracting atoms  with the same distances, ${\bf d}$ between neighboring atoms and denote by $N_e$ the number of electrons in each atom. Furthermore, we assume that the origin of coordinate coincides with the position of the first atom, so that the distance between the first and $a-$th atoms is  $(a-1){\bf d}$. Then the coordinate of the electron in $a-$th atom
is given by ${\bf R}_{a,e}=(a-1){\bf d}+{\bf r}_{a,e}$, where
 ${\bf r}_{a,e}$ is the distance between the electron and nucleus in the $a-$th atom ($a=1, 2, \cdots, N$).

Using these notation the potential for interaction of the chain with a pulse of an external electromagnetic field can be written as
\begin{eqnarray}
V(t)=
\sum_{a=1}^{N}\sum_{e=1}^{N_e}{\bf E}({\bf r}_{a,e}+(a-1){\bf d},t){\bf r}_{a,e} \;,
\label{iz2adfh}
\end{eqnarray}
To consider this potential as a sudden perturbation for a single atom the following condition should be obeyed:  $\tau \sim 1 / \alpha \ll \tau_a$.
However, for a chain of atoms with the length $ L = N | {\bf d} | $ the potential  $ V (t) $ can be considered as a sudden perturbation under the following conditions:
\begin{eqnarray}
\tau \sim 1/\alpha \ll \tau_a \sim 1,
\nonumber\\
N{\bf d}{\bf n}_0/c \ll \tau_a \sim 1.
\label{Ltau}
\end{eqnarray}

 We assume that each atom in the chain initially being the ground state, $\varphi _{0}(\{{\bf r}_{a,e}\}) $, with  $ \{{\bf r}_{a,e}\} $ being the set of coordinates of the atomic electrons.
Then for the wave function of the one-dimensional chain consisting of  $N$ noninteracting atoms with total  $NN_e$ electrons we have
\begin{eqnarray}
\Phi _{0}=\varphi _{0}(\{{\bf r}_{1,e}\}) \varphi _{0}(\{{\bf r}_{2,e}\})\cdots \varphi _{0}(\{{\bf r}_{N,e}\}).
\label{Phi}
\end{eqnarray}

In the following, we will be interested in the treatment of reradiation of the ultrashort electromagnetic pulse during the time its interaction with the above atomic chain.

Following the Ref. ~ \cite {Matv2003} the interaction of the atomic electrons with the electromagnetic field can be written as
\begin{eqnarray}
U = -\sum_{a=1}^{N}\sum_{e=1}^{N_e}\sum \limits_{{\bf k}, \sigma } \biggl(\frac{2\pi }
{\omega
  }\biggr) ^{\frac{1}{2}}{\bf u} _{{\bf k}\sigma}
  a_{{\bf k}\sigma}^{+}
 e^{-i{\bf k}{\bf R}_{a,e}}
{\bf \hat p}_{a}\;\;.
\label{U}
\end{eqnarray}
Here $ a_{{\bf k}\sigma}^{+} $  are the photon creation operators,
 $ \omega $, ${\bf k}$ and $\sigma,\; (\sigma =1,2)$ are the frequency, momentum and polarization of the created photon, respectively,
 ${\bf u}_{{\bf k}\sigma } $ are the unit vectors of polarization,  
while ${\bf \hat p}_{a,e}=\partial /\partial {\bf r}_{a,e}$ are the momentum operators of the atomic electrons.
Then the  amplitude for the  photon emission can be calculated in the first order of perturbation theory with respect to interaction $U$ ~ \cite {Matv2003}.

We want to find the photon radiation spectrum of on the solid angle $d\Omega_{\bf k}$ ,
drawn by the vector $ {\bf k}$. For this purpose we use an approach developed in the Ref.  ~\cite{Matv2003} which gives us
 (averaged over photon polarization vector and  summed over all the final states of atoms in the chain) the total spectrum
 of the photon radiation per unit of solid angle  $d\Omega_{\bf k}$:
\begin{eqnarray}
\frac{d^2W}{d\Omega_{\bf k}d\omega}= \frac{1}{(2\pi)^2}
\frac{1}{c^{3}\omega} \langle
 \Phi _{0} \mid \sum\limits_{a,e}  \sum\limits_{b,m} e^{-i{\bf k}
 ({\bf R}_{a,e}-{\bf R}_{b,m})}
 \times
\nonumber\\
\times
 [\frac{\partial
\widetilde V(\omega)}{\partial {\bf r}_{a,e}}{\bf n}]
[\frac{\partial
\widetilde V^{*}(\omega)}{\partial {\bf r}_{b,m}}
{\bf n}] \mid \Phi _{0} \rangle ,
 \label{iz10}
\end{eqnarray}
where ${\bf R}_{a,e}= (a-1){\bf d}+{\bf r}_{a,e}$ and   ${\bf R}_{b,e}=(b-1){\bf d}+{\bf r}_{b,e}$  are the coordinates of the electrons belonging
to $a-$th  $b-$th atoms, respectively, ${\bf n}={\bf k}/k$ is the unit vector along the direction of emitted photon and
$ \widetilde V(\omega)$ is the Fourier-transform of the potential $V(t)$, which is written as
\begin{eqnarray}
{\widetilde V}(\omega)= \int\limits_{-\infty}^{+\infty} V(t)
e^{i \omega t} dt= f_0(\omega)\sum\limits_{a,e}({\bf E}_0{\bf r}_{a,e})
\times
\nonumber\\
\times
exp\left(i\frac{\omega}
{\omega_0} {\bf k}_0((a-1){\bf d}+{\bf r}_{a.e})\right),
\label{izv}
\end{eqnarray}
with
\begin{eqnarray}
f_0(\omega)=
\frac{1}{2}
\Biggl\{exp\Biggl[-\frac{(\omega-\omega_0)^2}{4\alpha^2}\Biggr]+
\nonumber\\
+exp\left[-\frac{(\omega+\omega_0)^2}{4\alpha^2}\right] \Biggr\},
\label{izf}
\end{eqnarray}
and
\begin{eqnarray}
[\frac{\partial
\widetilde V(\omega)}{\partial {\bf r}_{a,e}}{\bf n}]=f_0(\omega)
exp\left\{i\frac{\omega}{\omega_0}{\bf k}_0((a-1){\bf d}+{\bf r}_{a,e})\right\}
\times
\nonumber\\
\times
\left([{\bf E}_0 {\bf n}] +i\frac{\omega}{\omega_0} ({\bf E}_0 {\bf r}_{a,e})
[{\bf k}_0 {\bf n}]\right).
\label{aa1}
\end{eqnarray}

Using Eqs. (\ref{izv}) -(\ref{aa1})the total radiation spectrum per unit od solid angle $d\Omega_{\bf k}$ for one pulse duration,  $ \tau$  can be written as
\begin{eqnarray}
\frac{d^2W}{d\Omega_{\bf k}d\omega}=  \frac{ \mid f_0(\omega) \mid ^{2}}{(2\pi)^2c^{3}\omega}
 \langle \Phi _{0} \mid \sum\limits_{a, e }\sum\limits_{b, m}
 e^{-i{\bf p}{\bf d}(a-b)}
 \times
\nonumber\\
\times
 e^{-i{\bf p}({\bf r}_{a, e }-{\bf r}_{b, m})}
 \Biggl\{[{\bf E}_0 {\bf n}]^2
+ i\frac{\omega}{\omega_0}
([{\bf E}_0 {\bf n}][{\bf k}_0 {\bf n}])
\times
\nonumber\\
\times
 {\bf E}_0({\bf r}_{a, e }-{\bf r}_{b, m})+
 \nonumber\\
 +({\bf E}_0{\bf r}_{a, e })({\bf E}_0{\bf r}_{b, m}) \frac{\omega^2}
 {\omega_0^2}[{\bf k}_0{\bf n}]^2\Biggr\} \mid \Phi _{0} \rangle.
 \label{iz10me}
\end{eqnarray}

It is clear that in this formula, the averages over the ground state, $ \Phi _ {0} $ are written in the  form
 $ \langle \Phi _ {0} \mid \sum \limits_ {e, m} f ({\bf r} _ {a, e}, {\bf r} _ {b, m}) \mid \Phi _ {0} \rangle $, and therefore do not depend on
 $ a $ and $ b $.

Separating in Eq. (\ref{iz10me}) summation for  $a=b$ and $a\not=b$ and denoting corresponding spectra by $d^2W_1/(d\Omega_{\bf k}d\omega)$
and $d^2W_2/(d\Omega_{\bf k}d\omega)$, respectively, we have
\begin{eqnarray}
\frac{d^2W}{d\Omega_{\bf k}d\omega}= \frac{d^2W_1}
{d\Omega_{\bf k}d\omega}
+\frac{d^2W_2}{d\Omega_{\bf k}d\omega}\;.
 \label{iz10h}
 \end{eqnarray}

In the following we will see that the term  $d^2W_1/(d\Omega_{\bf k}d\omega)$ describes incoherent (proportional to $ N $) rescattering ,
while the interference spectrum $d^2W_2/(d\Omega_{\bf k}d\omega)$, corresponds to appearing of characteristic diffraction maximums.

Let us apply the above formalism to the linear atomic chain consisting of $N$ non-interacting helium atoms.
If $\varphi_0({\bf r}_1, {\bf r}_2)$ is the ground  state wave function of helium atom, then the wave function of the chain,  $\Phi_0$
which appears in Eq. (\ref{iz10me}) can be written as
$$\Phi_0=\varphi_0({\bf r}_{1,1}, {\bf r}_{1,2})\varphi_0({\bf r}_{2,1}, {\bf r}_{2,2})\cdots\varphi_0({\bf r}_{N,1}, {\bf r}_{N,2}).$$

Furthermore, we write each ground state wave function of the helium atom,   $\varphi_0$ as the product of two  hydrogen like $1s$ wave functions
with an effective charge  $Z=2-5/16$. Then one can easily perform averaging over $\Phi_0$ in Eq. (\ref{iz10me})  and
the spectra can be written as
\begin{eqnarray}
\frac{d^2W_1}{d\Omega_{\bf k}d\omega}= \frac{ \mid f_0(\omega) \mid ^{2}}{(2\pi)^2c^{3}\omega}N \Biggl\{N_e G(\omega, {\bf n},{\bf n}_0)+
\nonumber\\
+ N_e(N_e-1)F(\omega, {\bf n},{\bf n}_0)\Biggr\} ,
 \label{w1helz}
 \end{eqnarray}
\begin{eqnarray}
\frac{d^2W_2}{d\Omega_{\bf k}d\omega}= \frac{ \mid f_0(\omega) \mid ^{2}}{(2\pi)^2c^{3}\omega}
N_e^2F(\omega, {\bf n},{\bf n}_0)g_N({\bf p}{\bf d})\;,
  \label{w2helz}
\end{eqnarray}
where
\begin{eqnarray}
G(\omega, {\bf n},{\bf n}_0)=[{\bf E}_0 {\bf n}]^2 +
 \frac{\omega^2} {c^2Z^2} E_0^2[{\bf n}_0{\bf n}]^2\;,
 \label{G}
\end{eqnarray}
\begin{eqnarray}
F(\omega, {\bf n},{\bf n}_0)=\left\{\frac{16}{[4+({\bf n}-{\bf n}_0)^2\omega^2/(cZ)^2]^2}\right\}^2
\times\nonumber\\\times
\Biggl\{[{\bf E}_0 {\bf n}] - \frac{\omega^2} {c^2Z^2}[{\bf n}_0{\bf n}]
\frac{4 ({\bf E}_0 ({\bf n}-{\bf n}_0))}{4+({\bf n}-{\bf n}_0)^2\omega^2/(cZ)^2}\Biggl\}^2\;,
  \label{F}
\end{eqnarray}
\begin{eqnarray}
g_N({\bf p}{\bf d})=\sum\limits_{a,b(a\not= b)} e^{{\bf p}{\bf d}(a-b)}=\frac{\sin^2({\bf p}{\bf d}N/2) }{\sin^2({\bf p}{\bf d}/2) }-N,
  \label{g}
\end{eqnarray}
\begin{eqnarray}
{\bf p}{\bf d}=\frac{\omega}{c}({\bf n}-{\bf n}_0){\bf d}.
  \label{tau}
\end{eqnarray}
Here ${\bf n}$ and ${\bf n}_0$ are the unit vectors along the directions of the emitted photon and incident ultrashort pulse, respectively,
 $\omega$ is the frequency of the emitted photon,   ${\bf d}$ is the vector directed along the chain such that $|{\bf d}|$ is the distance between the neigbouring atoms.
 Summing Eqs.  (\ref{w1helz}) and (\ref{w2helz}) we get the total radiation spectrum
\begin{eqnarray}
\frac{d^2W}{d\Omega_{\bf k}d\omega}=
 \frac{ \mid f_0(\omega) \mid ^{2}}{(2\pi)^2c^{3}\omega}
\Biggl\{N N_e\Biggl[G(\omega, {\bf n},{\bf n}_0)-
\nonumber\\
-F(\omega, {\bf n},{\bf n}_0)\Biggr]+
N_e^2 F(\omega, {\bf n},{\bf n}_0) \frac{\sin^2({\bf p}{\bf d}N/2) }{\sin^2({\bf p}{\bf d}/2) }\Biggr\}.
 \label{hg}
 \end{eqnarray}
This equation is valid for a chain with arbitrary $N$ provided the condition given by Eq. (\ref{Ltau})is obeyed.
For $Z=1$ и $N_e=1$ Eq. (\ref{hg}) describes reradiation of the chain of $N$ hydrogen atoms induced by the interaction of the chain with an ultrashort electromagnetic pulse. For  $N=1$ and $N_e=1$ it describes reradiation of single hydrogen atom while  $N=1$ and $N_e=2$ corresponds to the reradiation of single helium atom (see ~\cite{Matv2003}).

It is easy to see that near the zeros of its denominator,  second term in right hand side of  Eq. (\ref{hg})  behaves like $N^2$, while for
for  $N\gg 1$  it reproduces the  behavior which is usual for  of the angular distribution of the wave processes on diffraction gratings.
It should be noticed that the functions $G(\omega, {\bf n},{\bf n}_0)$ и $F(\omega, {\bf n},{\bf n}_0)$, contain (see Eqs.  (\ref{G}) and (\ref{F})), the frequency of radiated photon,  $\omega$ only in the form $\omega^2/(cZ)^2$.

Now let us consider behavior of of the spectrum given by Eq. (\ref{hg}) in the limit of small frequencies given by inequality  $\omega^2/(cZ)^2 \ll 1$. In this limit we have $G(\omega, {\bf n},{\bf n}_0)=F(\omega, {\bf n},{\bf n}_0)=[{\bf E}_0 {\bf n}]^2$ and Eq. (\ref{hg})  can be written as
\begin{eqnarray}
\frac{d^2W}{d\Omega_{\bf k}d\omega}=
\frac{ \mid f_0(\omega) \mid ^{2}}{(2\pi)^2c^{3}\omega}
N_e^2 [{\bf E}_0 {\bf n}]^2 \frac{\sin^2({\bf p}{\bf d}N/2) }{\sin^2({\bf p}{\bf d} /2) }.
 \label{hgsmall}
 \end{eqnarray}
The contribution from the interference effects can be characterized by the quantity
\begin{eqnarray}
\frac{\frac{d^2W}{d\Omega_{\bf k}d\omega}}
{\frac{d^2W_1}{d\Omega_{\bf k}d\omega}}
=\frac{1}{N}\frac{\sin^2({\bf p}{\bf d}N/2) }{\sin^2({\bf p}{\bf d} /2) }\;.
 \label{delta}
\end{eqnarray}
 In this formula zeros of denominator in the  right hand side correspond to maximum values of the angular distribution. In the limit  $N \gg 1$
 for ${\bf p}{\bf d}/2 =\pi n+ \epsilon$ ($n$ is integer and, $\epsilon$ is a small quantity), we have an asymptotic relation \cite{LandauII}
\begin{eqnarray}
\lim_{N\to \infty}\frac{1}{N}\frac{\sin^2({\bf p}{\bf d} N/2) }{\sin^2({\bf p}{\bf d} /2) }=
\lim_{N\to \infty}\frac{1}{N}\frac{\sin^2(\epsilon N) }{\epsilon^2}
= \pi \delta (\epsilon).
\end{eqnarray}

It is easy to see that in Eq. (\ref{hg}) near the above mentioned zeros of denominator second term behaves like $(N_eN)^2$  and corresponds to coherent radiation of photon by all electrons in the chain.  As it can been from Eq.(\ref{hg})for   $\omega^2/(cZ)^2 \to \infty$ (or, equivalently, $\omega^2/(cZ)^2 \gg NN_e$) the coherent part of the spectrum is much smaller compared to incoherent part (which is proportional to $NN_e$). Therefore increasing of $\omega$ leads to decreasing of the contribution by coherent radiation.

Eq.(\ref{hg}) can be utilized for qualitative description of reradiation process caused by the interaction with spatially non-uniform  ultrashort pulse for arbitrary number of atomic electrons ($N_e> 2$) and atoms in the chain.
It follows from Eq.(\ref{iz10me}) that the dependence  of the total spectrum on  $N$ and $N_e$ has the same character as that in Eq.(\ref{hg}). The only difference comes from the functions  $G(\omega, {\bf n},{\bf n}_0)$ and $F(\omega, {\bf n},{\bf n}_0)$ which are calculated as the averages over the ground state wave functions of some atoms in the chain. In the above treatment we considered atoms in the chain as noninteracting to each other ignoring by electron-electron correlations. However, screening of nucleus charge by inner electrons has been taken into account using the effective charge in Eqs.  (\ref{w1helz}) and (\ref{w2helz}).  Finally we note that the above treatment of reradiation process induced in linear atomic chain by its interaction with a spatially non-uniform ultrashort pulse can be extended to the cases of higher dimensional (e.g., graphene, three-dimensional crystall lattices etc) atomic chains, too. Also, it is possible to include into consideration thermal vibrations of the atoms in the chain. Such a study is currently on progress.


\begin{thebibliography}{150}
\bibitem{Kauli}  John M. Cowley  {\it Diffraction Physics} (North-Holland, Amsterdam), 1975.
\bibitem{Moshammer1997}  R. Moshammer,  W. Schmitt,  J. Ullrich et al. Phys. Rev. Lett. {\bf 79},
3621 (1997).
\bibitem{Agostini2004}  P. Agostini Rep. Prog. Phys. {\bf 67}. 813 (2004).
\bibitem{Corkit2007} P. В. Corkit,  F. Krausz Nature Phys. {\bf 3}, 381 (2007).
\bibitem{Krausz2009} F. Kraus ,  M.Ivanov Rev. Mod. Phys. {\bf 81}, 163 (2009).
\bibitem{Jeltikov2009} A.M. Zheltikov Phys.Usp. {\bf 54} 29 (2011).
\bibitem{Astapenko2011} V.A. Astapenko JETP {\bf 112} 193 (2011).
\bibitem{Golovinkii2006} P.A. Golovinkii, E.M. Mikhailov,  Laser Phys. Lett. {\it 3}, 259, (2006).
\bibitem{Astapenko2010} V.A. Astapenko  Phys. Lett., {\bf A374} 1585 (2010).
\bibitem{Matv2003} V. I. Matveev, JETP {\bf 97} , 915 (2003).
\bibitem{Eseev2011} M. K. Eseev, V. I. Matveev and V. M. Yulkova, Optics and Spectroscopy, , {\bf  111} 33 (2011).
\bibitem{DykhneYudin1978} A. M. Dykhne and G. L. Yudin, Sov. Phys.–Usp. { \bf 21} 549 (1978).
\bibitem{LandauII} L. D. Landau and E. M. Lifshitz, Course of Theoretical Physics, Vol. 2: The Classical Theory of Fields (Pergamon, Oxford, 1975).

\end{thebibliography}
\end{document}